\documentclass[twocolumn,showpacs,preprintnumbers,amsmath,amssymb,prl]{revtex4}

\usepackage{graphicx}
\usepackage{dcolumn}
\usepackage{bm}
\usepackage{tabularx}
\usepackage{amsmath}

\begin{document}

\title{Quantum criticality and nodal superconductivity in the
FeAs-based superconductor KFe$_2$As$_2$ }

\author{J. K. Dong,$^1$ S. Y. Zhou,$^1$ T. Y. Guan,$^1$ H. Zhang,$^1$ Y. F. Dai,$^1$ X. Qiu,$^1$ X. F. Wang,$^2$ Y. He,$^2$ X. H. Chen,$^2$ S. Y. Li$^{1,*}$}

\affiliation{$^1$Department of Physics, Surface Physics Laboratory (National Key Laboratory), and Laboratory of Advanced Materials, Fudan University, Shanghai 200433, China\\
$^2$Hefei National Laboratory for Physical Science at Microscale and
Department of Physics, University of Science and Technology of
China, Hefei, Anhui 230026, China}

\date{\today}

\begin{abstract}
The in-plane resistivity $\rho$ and thermal conductivity $\kappa$ of
FeAs-based superconductor KFe$_2$As$_2$ single crystal were measured
down to 50 mK. We observe non-Fermi-liquid behavior $\rho(T) \sim
T^{1.5}$ at $H_{c_2}$ = 5 T, and the development of a Fermi liquid
state with $\rho(T) \sim T^2$ when further increasing field. This
suggests a field-induced quantum critical point, occurring at the
superconducting upper critical field $H_{c_2}$. In zero field there
is a large residual linear term $\kappa_0/T$, and the field
dependence of $\kappa_0/T$ mimics that in $d$-wave cuprate
superconductors. This indicates that the superconducting gaps in
KFe$_2$As$_2$ have nodes, likely $d$-wave symmetry. Such a nodal
superconductivity is attributed to the antiferromagnetic spin
fluctuations near the quantum critical point.
\end{abstract}


\maketitle

When superconductivity emerges with the suppression of magnetism,
for example in heavy-fermion and high-temperature cuprate
superconductors, the spin fluctuations associated with a magnetic
quantum critical point is usually considered as the pairing glue.
This is also the case for the recently discovered FeAs-based
high-temperature superconductors \cite{Kamihara,Chen,Rotter,Liu}.
The parent compounds of the FeAs-based superconductors, for example
LaFeAsO and BaFe$_2$As$_2$, are not superconducting and manifest
antiferromagnetic (AF) order \cite{Cruz,QHuang}. With electron or
hole doping, the AF order is suppressed and superconductivity
emerges \cite{Kamihara,Chen,Rotter,Liu}.

Spin fluctuations usually result in superconducting gaps with nodes,
but it can also give nodeless superconducting gaps through interband
interaction, termed $s_\pm$-wave \cite{Mazin1,KKuroki}. In the
$s_\pm$-wave pairing theory for FeAs-based superconductors
\cite{Mazin1}, the interband interaction happens between the hole
pockets at $\Gamma$ point and the electron pockets at M point, via
the antiferromagnetic spin fluctuations (AFSF) with wave vector
$Q_{AF}$ = ($\pi$,$\pi$). This gives full superconducting gaps on
both electron and hole pockets, but with the opposite signs of the
order parameters between the two \cite{Mazin1}. While there are
accumulating experimental and theoretical works in favor of nodeless
gaps, conclusive phase-sensitive experiments are needed to confirm
the $s_\pm$-wave pairing in FeAs-based superconductors
\cite{Mazin2}. Moreover, in LaFePO and
BaFe$_2$(As$_{1-x}$P$_x$)$_2$, two compounds containing phosphorus,
there are evidences for nodal superconductivity
\cite{Fletcher,Hicks,Hashimoto,Nakai}. Therefore, the pairing
symmetry and superconducting mechanism in iron pnictides are still
far from consensus.

Recently, the ARPES experiments on the extremely hole-doped
KFe$_2$As$_2$ ($T_c$ = 3 K) showed that the electron pockets at M
point completely disappear due to the hole doping \cite{TSato}. This
result immediately raises a very important question: what is the
superconducting state in heavily overdoped FeAs-based
superconductors where the interband interaction is suppressed? While
ARPES experiment was unable to study the superconducting gap
structure in KFe$_2$As$_2$ due to its low $T_c$, the low-temperature
thermal conductivity technique is particularly useful for studying
exotic superconductors with low $T_c$ \cite{Shakeripour}.

In this Letter, we report the demonstration of a clear field-induced
antiferromagnetic quantum critical point and nodal superconductivity
in KFe$_2$As$_2$ by resistivity and thermal conductivity
measurements. Our findings not only confirm the
spin-fluctuation-mediated pairing mechanism, but also clarify the
pairing symmetry when interband interaction is suppressed in heavily
overdoped regime, thus complete our understanding of the
superconducting state in the FeAs-based superconductors.

Single crystals of KFe$_2$As$_2$ were grown by self-flux method
\cite{XFWang}. The dc magnetic susceptibility was measured by a
SQUID (Quantum Design). The sample was cleaved to a rectangular
shape of dimensions 1.5 $\times$ 1.0 mm$^2$
 in the $ab$-plane, with 40 $\mu$m thickness
along the $c$-axis. Contacts were made directly on the sample
surfaces with silver paint, which were used for both resistivity and
thermal conductivity measurements. To avoid degradation, the sample
was exposure in air less than 2 hours. The contacts are metallic
with typical resistance 100 m$\Omega$ at 1.5 K. In-plane thermal
conductivity was measured in a dilution refrigerator, using a
standard four-wire steady-state method with two RuO$_2$ chip
thermometers, calibrated {\it in situ} against a reference RuO$_2$
thermometer. Magnetic fields were applied along the $c$-axis and
perpendicular to the heat current. To ensure a homogeneous field
distribution in the sample, all fields were applied at temperature
above $T_c$.

\begin{figure}
\includegraphics[clip,width=6cm]{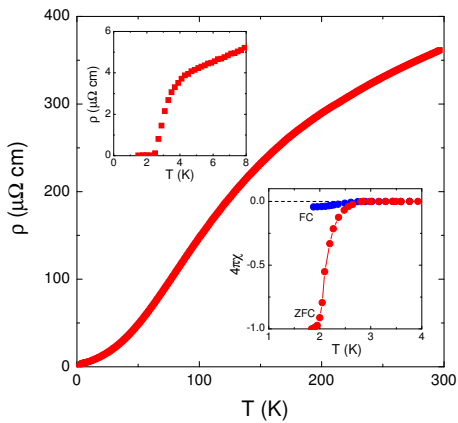}
\caption{(Color online). In-plane resistivity of KFe$_2$As$_2$
single crystal. The residual resistivity ratio is $\rho$(297
K)/$\rho$(5 K) = 86. Upper inset: the resistive transition at low
temperature. Lower inset: the dc magnetic susceptibility at $H$ = 10
Oe, with both zero field cooled (ZFC) and field cooled (FC)
measuring conditions. }
\end{figure}

\begin{figure}
\includegraphics[clip,width=5cm]{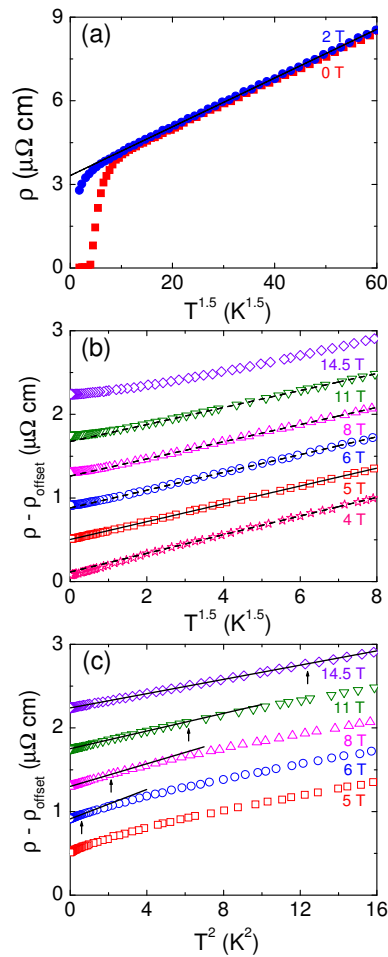}
\caption{(Color online). (a) Low-temprature resistivity of
KFe$_2$As$_2$ single crystal in $H$ = 0 and 2 T plotted as $\rho$ vs
$T^{1.5}$ . The solid line is a fit of the $H$ = 2 T data between 4
and 16 K to $\rho$ = $\rho_{0}$ + $AT^{1.5}$ , which gives residual
resistivity $\rho_{0}$ = 3.32 $\mu$$\Omega$ cm. (b) $\rho$ vs
$T^{1.5}$ for $H$ = 4, 5, 6, 8, 11, and 14.5 T (data sets are offset
for clarity). The solid line is a fit of the $H$ = 5 T data between
50 mK and 4 K. The dash lines are guides to the eye for the
deviation from the $T^{1.5}$ dependence. (c) $\rho$ vs $T^2$ for $H$
= 5, 6, 8, 11, and 14.5 T (data sets are offset for clarity). The
solid lines are fits to $\rho$ = $\rho_{0}$ + $AT^{2}$. The arrows
indicate the upper limit of the temperature range of $T^2$ behavior.
}
\end{figure}

Fig. 1 shows the in-plane resistivity of our KFe$_2$As$_2$ single
crystal. The residual resistivity ratio (RRR) $\rho$(297 K)/$\rho$(5
K) = 86, which is very close to that reported previously, RRR = 87
\cite{TTerashima}. The upper and lower insets of Fig. 1 show the
resistive and magnetic superconducting transitions at low
temperature. The middle point of the resistivity transition is at
$T_c$ = 3.0 K.

In Fig. 2a, the resistivity in $H$ = 0 and 2 T are plotted as $\rho$
vs $T^{1.5}$. It is found that $\rho$ obeys $T^{1.5}$ dependence
nicely above $T_c$, up to about 20 K. Previously, Terashima {\it et
al.} claimed that $\rho$ exhibits a $T^2$ dependence below $\sim$ 45
K \cite{TTerashima}. However, we note that their $T^2$ fitting does
not look good at low temperature. To elucidate how low the
$T^{1.5}$-dependent $\rho(T)$ can go, we measure the resistivity
down to 50 mK in a dilution refrigerator and in higher magnetic
fields. Fig. 2b plots $\rho$ vs $T^{1.5}$ for $H$ = 4, 5, 6, 8, 11,
and 14.5 T. The downward deviation of $\rho$ from $T^{1.5}$
dependence below 1.6 K in $H$ = 4 T is attributed to the onset of
superconductivity. In $H$ = 5 T, we find a perfect
$T^{1.5}$-dependent resistivity down to 50 mK. In $H >$ 5 T, there
is an upward deviation of $\rho$ from the $T^{1.5}$ dependence at
low temperature. The data of $H$ = 5, 6, 8, 11, and 14.5 T were
re-plotted as $\rho$ vs $T^{2}$ in Fig. 2c. It is clearly seen that
a Fermi liquid behavior of resistivity, $\rho$ $\sim$ $T^{2}$,
develops with increasing field.

Based on our experimental determined ranges of $T^{2}$ behavior at
low temperatures, we have constructed a phase diagram of the $T - H$
plane (Fig. 3). The inset of Fig. 3 plots the field dependence of
the coefficient $A$ of the $T^{2}$ term, which tends to diverge
towards $H_{c2}$ = 5 T. Such a phase diagram of KFe$_2$As$_2$ is
strikingly similar to that of the heavy-fermion superconductor
CeCoIn$_5$ with $T_c$ = 2.3 K, in which a field-induced AF quantum
critical point (QCP) is located at $H_{c2}$
\cite{Paglione1,Bianchi,Paglione2}.

Near an AF QCP, the scattering of electrons by AFSF usually leads to
non-Fermi-liquid behavior of resistivity, $\rho$ $\sim$ $T^{1.5}$ in
3D system and $\rho$ $\sim$ $T$ in 2D system \cite{Stewart}. The
observation of $\rho$ $\sim$ $T^{1.5}$ at $H_{c2}$ in quasi-2D
CeCoIn$_5$ is explained by the similar character of the magnetic
fluctuations in the CeIn$_3$ planes of CeCoIn$_5$ and in bulk 3D
CeIn$_3$ itself \cite{Paglione2}. As for KFe$_2$As$_2$ in this
study, the $T^{1.5}$ dependence of $\rho$ at $H_{c2}$ indicates that
the magnetic fluctuations in KFe$_2$As$_2$ also have some 3D
character, which may need further investigation.

The similarity between the $T - H$ phase diagrams of KFe$_2$As$_2$
and CeCoIn$_5$ suggests that there is also a field-induced AF QCP at
$H_{c2}$ in KFe$_2$As$_2$. To our knowledge, this is the first time
to demonstrate an AF QCP in FeAs-based superconductors down to low
temperature.

A field-induced AF QCP at $H_{c2}$ is very unusual, since it
indicates that the superconducting and magnetic orders are tightly
coupled \cite{Kenzelmann}. This may be easy to understand in
CeCoIn$_5$, since there is a closely related compound CeRhIn$_5$
which is an ambient pressure antiferromagnet with the N{\'e}el
temperature $T_N$ = 3.8 K. For KFe$_2$As$_2$, this is highly
unexpected, in the sense that KFe$_2$As$_2$ is far away from the AF
parent compound BaFe$_2$As$_2$ in Ba$_{1-x}$K$_x$Fe$_2$As$_2$
system. However, our finding is strongly supported by recent nuclear
magnetic resonance (NMR) experiments on KFe$_2$As$_2$ single
crystal, which did claim the existence of strong AFSF
\cite{SWZhang}.

\begin{figure}
\includegraphics[clip,width=6cm]{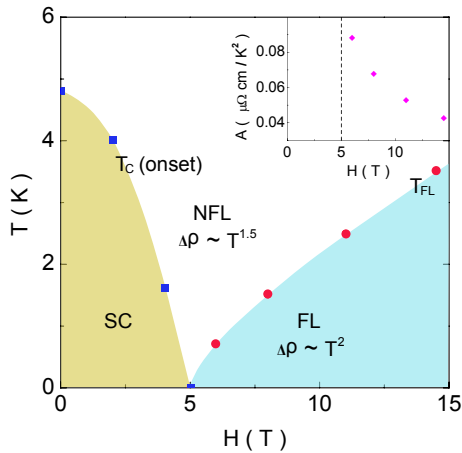}
\caption{(Color online). The $T - H$ phase diagram determined from
the resistivity measurements. The $T_c$(onset) is defined at the
temperature where $\rho$ deviates downwards from the $T^{1.5}$
dependence. The $T_{FL}$ is defined as the upper limit of the
temperature range of $T^{2}$ dependence Fermi liquid behavior. This
shows a clear field-induced quantum critical point located at
$H_{c2}$ = 5 T. The inset shows the field dependence of the
coefficient $A$ of $\rho = \rho_0 + AT^2$, which tends to diverge
towards $H_{c2}$ = 5 T. }
\end{figure}

\begin{figure}
\includegraphics[clip,width=5.5cm]{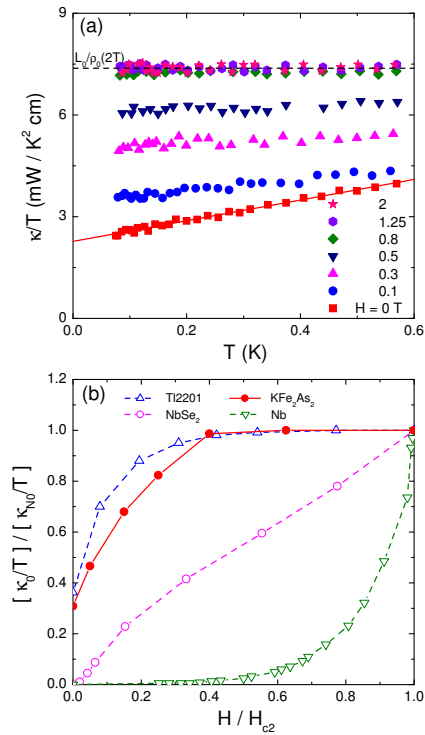}
\caption{(Color online). (a) Low-temperature in-plane thermal
conductivity of KFe$_2$As$_2$ in magnetic fields applied along the
c-axis ($H$ = 0, 0.1, 0.3, 0.5, 0.8, 1.25 and 2 T). The solid line
is $\kappa/T = a + bT$ fit to the $H$ = 0 T data. The dash line is
the normal state Wiedemann-Franz law expectation $L_0$/$\rho_0$(2
T), with $L_0$ the Lorenz number 2.45 $\times$ 10$^{-8}$
W$\Omega$K$^{-2}$ and $\rho_0$(2 T) = 3.32 $\mu\Omega$ cm. (b)
Normalized residual linear term $\kappa_0/T$ of KFe$_2$As$_2$ as a
function of $H/H_{c2}$. Similar data of the clean $s$-wave
superconductor Nb \cite{Lowell}, the multi-band $s$-wave
superconductor NbSe$_2$ \cite{Boaknin}, and an overdoped sample of
the $d$-wave superconductor Tl-2201 \cite{Proust} are also shown for
comparison. The behavior of $\kappa_0(H)/T$ in KFe$_2$As$_2$ clearly
mimics that in Tl-2201.}
\end{figure}

Having demonstrated the field-induced AF QCP in KFe$_2$As$_2$, we
continue to investigate its superconducting gap structure. Fig. 4a
shows the temperature dependence of the in-plane thermal
conductivity for KFe$_2$As$_2$ in $H$ = 0, 0.1, 0.3, 0.5, 0.8, 1.25,
and 2 T magnetic fields, plotted as $\kappa/T$ vs $T$. All the
curves are roughly linear, as previously observed in
BaFe$_{1.9}$Ni$_{0.1}$As$_2$ and overdoped BaFe$_{2-x}$Co$_x$As$_2$
single crystals \cite{LDing,Tanatar,JKDong}. Therefore we fit the
data to $\kappa/T$ = $a + bT^{\alpha-1}$ with $\alpha$ fixed to 2.
The two terms $aT$ and $bT^{\alpha}$ represent contributions from
electrons and phonons, respectively. Here we only focus on the
electronic term.

In zero field, the fitting gives a residual linear term $\kappa_0/T$
= 2.27 $\pm$ 0.02 mW K$^{-2}$ cm$^{-1}$. Such a large $\kappa_0/T$
in KFe$_2$As$_2$ is really very surprising, since previous thermal
conductivity studies of FeAs-based superconductors (without
phosphorus), including Ba$_{0.75}$K$_{0.25}$Fe$_2$As$_2$, have given
negligible $\kappa_0/T$ in $H$ = 0
\cite{LDing,Tanatar,JKDong,XGLuo}. From Fig. 4a, a very small field
$H$ = 0.1 T has significantly increased the $\kappa/T$. Above $H$ =
0.8 T, $\kappa/T$ tends to saturate. In $H$ = 1.25 and 2 T,
$\kappa_0/T$ = 7.39 $\pm$ 0.03 and 7.36 $\pm$ 0.04 mW K$^{-2}$
cm$^{-1}$ were obtained from the fittings, respectively. Both values
meet the expected normal state Wiedemann-Franz law expectation
$L_0$/$\rho_0$(2 T) = 7.38 mW K$^{-2}$ cm$^{-1}$, within
experimental error bar. We take $H$ = 2 T as the bulk $H_{c2}$ of
KFe$_2$As$_2$, despite that the resistive transition is not
completely suppressed until $H_{c2}$(onset) = 5 T. To choose a
slightly different bulk $H_{c2}$ does not affect our discussions
below.

In Fig. 4b, the normalized $\kappa_0/T$ of KFe$_2$As$_2$ is plotted
as a function of $H/H_{c2}$, together with the clean $s$-wave
superconductor Nb \cite{Lowell}, the multi-band $s$-wave
superconductor NbSe$_2$ \cite{Boaknin}, and an overdoped sample of
the $d$-wave superconductor Tl-2201 \cite{Proust}. For
KFe$_2$As$_2$, the large $\kappa_0/T$ in $H$ = 0 and the rapid
increase of $\kappa_0/T$ at low field mimic the behavior of Tl-2201,
and provide clear evidences for unconventional superconducting gap
with nodes \cite{Shakeripour}. We note that recent $^{75}$As nuclear
quadrupole resonance (NQR) and specific heat measurements on
KFe$_2$As$_2$ polycrystals also suggested multiple nodal gaps
\cite{Fukazawa}.

The nodal gap in extremely hole-doped KFe$_2$As$_2$ is distinctly
different from the nodeless gaps in FeAs-based superconductors at
other doping \cite{LDing,Tanatar,JKDong,XGLuo,PRichard,KTerashima}.
In optimally hole-doped Ba$_{0.6}$K$_{0.4}$Fe$_2$As$_2$ and
electron-doped BaFe$_{1.85}$Co$_{0.15}$As$_2$, the observations of
nearly nested Fermi-surface pockets and nodeless gaps suggest that
the interband interaction may play a crucial role in superconducting
pairing \cite{PRichard,KTerashima}, thus support the $s_{\pm}$-wave
pairing mechanism in FeAs-based superconductors \cite{Mazin1}.
However, in KFe$_2$As$_2$, the electron pockets at M point
completely disappear due to hole doping \cite{TSato}. Although four
new small hole ($\epsilon$) pockets are found around M point, the
interband interaction is nevertheless suppressed, therefore the
$s_{\pm}$-wave pairing mechanism does not work in KFe$_2$As$_2$
\cite{TSato}. Having known that there exist strong AFSF near the AF
QCP, one can be sure that the nodal superconductivity in
KFe$_2$As$_2$ results from intraband pairing via AFSF. Usually, the
pairing mediated by AFSF has $d$-wave symmetry, as in CeCoIn$_5$
\cite{Vorontsov}. Because of the great similarity between the $T -
H$ phase diagrams of KFe$_2$As$_2$ and CeCoIn$_5$, and the similar
behavior of $\kappa_0(H)/T$ between KFe$_2$As$_2$ and Tl-2201, the
nodal gap in KFe$_2$As$_2$ is very likely also $d$-wave.

We note that there is an electron-hole asymmetry in the phase
diagram of FeAs-based superconductors at the heavily overdoped
regime. For the heavily electron-doped BaFe$_{1.7}$Co$_{0.3}$As$_2$,
the hole pocket at $\Gamma$ point completely disappears
\cite{YSekiba}, and the interband interaction is also suppressed.
But it turns out that BaFe$_{1.7}$Co$_{0.3}$As$_2$ is not
superconducting \cite{YSekiba}. This electron-hole asymmetry may be
explained by the different strength of AFSF on electron- and
hole-doped sides measured by NMR \cite{SWZhang,FLNing}. In
BaFe$_{2-x}$Co$_x$As$_2$, the spin fluctuations are completely
suppressed at x $>$ 0.3 \cite{FLNing}, while strong AF spin
fluctuations were found in KFe$_2$As$_2$ \cite{SWZhang}.

Finally, it is worth to point out that the possible nodal
superconductivity in LaFePO and BaFe$_2$(As$_{1-x}$P$_x$)$_2$
\cite{Fletcher,Hicks,Hashimoto,Nakai} may be different from that in
KFe$_2$As$_2$. The main reason is that the isovalent substitution of
P for As only slightly modifies the Fermi surface \cite{Hashimoto},
therefore the Fermi-surface nesting and interband interaction still
exist in these two compounds. To get nodal superconductivity in
LaFePO and BaFe$_2$(As$_{1-x}$P$_x$)$_2$, one needs to consider the
competition between $s_{\pm}$-wave and $d$-wave superconducting
states even more carefully.

In summary, we have measured the resistivity and thermal
conductivity of extremely hole-doped KFe$_2$As$_2$ single crystal
down to 50 mK. A field-induced AF QCP is demonstrated by the
observation of $\rho \sim T^{1.5}$ at $H_{c2}$ = 5 T, and the
development of $\rho \sim T^2$ Fermi liquid behavior at $H >$ 5 T.
Furthermore, the large $\kappa_0/T$ at zero field and a rapid
increase of $\kappa_0(H)/T$ at low field give strong evidences for
nodal superconductivity in KFe$_2$As$_2$. Such a nodal
superconductivity, very likely $d$-wave, naturally results from the
intraband pairing via AFSF near the AF QCP. Our results are
consistent with the suppression of interband interaction in
KFe$_2$As$_2$, as revealed by ARPES experiments.

We thank Y. Chen, D. L. Feng, and Y. Y. Wang for discussions. This
work is supported by the Natural Science Foundation of China, the
Ministry of Science and Technology of China (National Basic Research
Program No:2009CB929203), Program for New Century Excellent Talents
in University, and STCSM of China (No:08dj1400200 and 08PJ1402100). \\

$^*$ E-mail: shiyan$\_$li@fudan.edu.cn

\end{document}